\documentclass[twocolumn]{aastex63}
\title{A2018V3-REU-Paper}

\newcommand\aastex{AAS\TeX}

\usepackage{xcolor}
\usepackage{url}
\usepackage{amsmath}

\submitjournal{The Planetary Science Journal}

\shorttitle{\aastex\ Manx Candidate A/2018 V3}
\shortauthors{Piro et al.}

\begin{document}

\title{Characterizing the Manx Candidate A/2018 V3}

\correspondingauthor{Karen J. Meech}
\email{meech@ifa.hawaii.edu}

\author[0000-0001-6160-050X]{Caroline Piro}
\affiliation{Institute for Astronomy, University of Hawaii, 2680 Woodlawn Drive, Honolulu, HI 96822, USA}

\author[0000-0002-2058-5670]{Karen J. Meech}
\affiliation{Institute for Astronomy, University of Hawaii, 2680 Woodlawn Drive, Honolulu, HI 96822, USA}

\author[0000-0002-0406-8518]{Erica Bufanda}
\affiliation{Institute for Astronomy, University of Hawaii, 2680 Woodlawn Drive, Honolulu, HI 96822, USA}

\author[0000-0002-4734-8878]{Jan T. Kleyna}
\affiliation{Institute for Astronomy, University of Hawaii, 2680 Woodlawn Drive, Honolulu, HI 96822, USA}

\author[0000-0002-2021-1863]{Jacqueline V. Keane}
\affiliation{Institute for Astronomy, University of Hawaii, 2680 Woodlawn Drive, Honolulu, HI 96822, USA}

\author[0000-0001-6952-9349]{Olivier Hainaut}
\affiliation{European Southern Observatory, Karl-Schwarzschild-Strasse 2, D-85748 Garching bei M\"unchen, Germany}

\author[0000-0001-7895-8209]{Marco Micheli}
\affiliation{ESA NEO Coordination Centre, Largo Galileo Galilei, 1, 00044 Frascati (RM), Italy}
\affiliation{INAF - Osservatorio Astronomico di Roma, Via Frascati 33, 00040 Monte Porzio Catone (RM), Italy}

\author[0000-0003-1790-5749]{James Bauer}
\affiliation{University of Maryland, Dept. of Astronomy, College Park, MD 20742-2421 USA}

\author[0000-0002-7034-148X]{Larry Denneau}
\affiliation{Institute for Astronomy, University of Hawaii, 2680 Woodlawn Drive, Honolulu, HI 96822, USA}

\author[0000-0002-0439-9341]{Robert Weryk}
\affiliation{Institute for Astronomy, University of Hawaii, 2680 Woodlawn Drive, Honolulu, HI 96822, USA}

\author[0000-0003-0174-3829]{Bhuwan C. Bhatt}
\affiliation{Indian Institute of Astrophys., II Block, Koramangala, Bangalore 560 034, India}

\author[0000-0001-9701-4625]{Devendra K. Sahu}
\affiliation{Indian Institute of Astrophys., II Block, Koramangala, Bangalore 560 034, India}

\author[0000-0002-1341-0952]{Richard Wainscoat}
\affiliation{Institute for Astronomy, University of Hawaii, 2680 Woodlawn Drive, Honolulu, HI 96822, USA}

\begin{abstract}
Manx objects approach the inner solar system on long-period comet (LPC) orbits with the consequent high inbound velocities, but unlike comets, Manxes display very little to no activity even near perihelion.  This suggests that they may have formed in circumstances different from typical LPCs; moreover, this lack of significant activity also renders them difficult to detect at large distances.  Thus, analyzing their physical properties can help constrain models of solar system formation as well as sharpen detection methods for those classified as NEOs.  Here, we focus on the Manx candidate A/2018 V3 as part of a larger effort to characterize Manxes as a whole.  This particular object was observed to be inactive even at its perihelion at $q$ = 1.34 au in 2019 September.  Its spectral reflectivity is consistent with typical organic-rich comet surfaces with colors of $g'-r'= 0.67\pm0.02$, $r'-i' = 0.26\pm0.02$, and $r'-z' = 0.45\pm0.02$, corresponding to a spectral reflectivity slope of $10.6\pm 0.9$ \%/100nm. A least-squares fit of our constructed light curve to the observational data yields an average nucleus radius of $\approx$2 km assuming an albedo of 0.04.  This is consistent with the value measured from NEOWISE.  A surface brightness analysis for data taken 2020 July 13 indicated possible low activity ($\lesssim0.68$ g $\rm s^{-1}$), but not enough to lift optically significant amounts of dust.  Finally, we discuss Manxes as a constraint on solar system dynamical models as well as their implications for planetary defense.

\end{abstract}


\section{Introduction} 
\label{sec:intro}

Many models of early solar system dynamics include periods where the giant planets may have shifted about radially before settling into their present positions \citep{gomes2005, walsh2011, raymond2017}.  In the process, small solar system objects were scattered from their original locations.  However, the models vary in both the magnitude and timing of planetary migration necessary to result in the solar system arrangement we see today.  In addition, dynamical formation models disagree on how much material may have originally been present in the inner solar system and what percentage may have been ejected to the Oort cloud during the periods of giant planet migration \citep{raymond2017}.

Manx objects may help us distinguish among these solar system models. Manxes are small solar system bodies that blur the line between asteroids and comets.  These objects follow long-period comet (LPC) orbits but exhibit very little, if any, of the cometary activity expected from an LPC. The term ``Manx'' was coined by \citet{meech2014} after the breed of tailless cat to describe the comet C/2013 P2 (PANSTARRS).  Despite a typical LPC orbit, C/2013 P2 displayed orders of magnitude less activity than typical LPCs, much less than even low-activity short period comets.  Even near its perihelion at 2.83 au, C/2013 P2 exhibited only a minimal tail.  A number of other low-activity objects on LPC orbits have since been observed \citep{stephens2017,bufanda2020} or identified as such in literature searches \citep{weissman1997,sekiguchi2018}.

Most intriguingly, the Manx object C/2014 S3 (PANSTARRS) was observed to have a spectral reflectivity with the $1.0~\micron$~absorption typical of inner solar system rocky asteroids and inconsistent with an inactive comet \citep{meech2016}.  The stony S-type asteroids dominate the population of the inner asteroid belt \citep{gaffey1993, raymond2017}. This surface type had never been seen on objects on an LPC orbit, although dynamical models predict that inner solar system material can be scattered to the Oort cloud.  Even more striking, the minerals that form the $1.0~\micron$~absorption characteristic of S-types cannot form in the presence of water, yet the activity seen in both C/2013 P2 and C/2014 S3 was consistent with the presence of water ice sublimation \citep{meech2016}.  It is hoped that the fraction of Manx objects that display S-type spectral reflectivity, among a sample of Manxes, can be extrapolated to the Oort cloud as a whole and thus help constrain the models of early solar system dynamics \citep{meech2016,shannon2019}.  

We have studied a Manx discovered by the Panoramic Survey Telescope and Rapid Response System (Pan-STARRS2) on Haleakal\={a}, Maui \citep{hoegner2018}.  This object, later designated A/2018 V3, was first detected at $r=3.99$ au on 2018 November 2 with $g$-band magnitude of 21.9, suggesting a km-scale nucleus.  Further observations over the course of the month allowed a good orbit to be determined.  It is on a long-period comet orbit ($a = 121.59$ au; $e = 0.989$; $q = 1.34$ au; $Q = 241.8$ au; $i = 165\degr$; $P = 1341$ yr at the time of perihelion\footnote{JPL Small-Body Database Browser: \url{https://ssd.jpl.nasa.gov/sbdb.cgi}}), but no activity was observed even at distances where significant activity would be expected for a typical LPC.  Thus, under the criteria set out by \citet{meech2016}, A/2018 V3 currently qualifies as a Manx object.

With a perihelion of $q = 1.34$ au, A/2018 V3 falls just outside the definition of a Near-Earth Object (NEO).  NEOs are typically defined as any comet or asteroid with a perihelion $q < 1.3$ au\footnote{NASA/JPL/CNEOS definitions of NEO groups: \url{https://cneos.jpl.nasa.gov/about/neo_groups.html}}, but this particular object piqued our interest when it passed perihelion and remained inactive.  LPCs can pose significant hazards to Earth because of their high velocity \citep{meech2017}; Manx objects pose an even greater hazard.  Their lack of activity, difficulty of detection, and LPC orbits could all result in very little warning time for a possible impact \citep{chodas1996,weissman1997a,nuth2018}.  

A/2018 V3 passed perihelion in 2019 September, but given the nature of Manxes, there are likely many other such inconspicuous objects traveling through the inner solar system.  Characterizing A/2018 V3 will allow us to investigate the role of Manxes in the overall context of solar system dynamics and will sharpen our ability to detect more of these small, dark solar system objects that still carry sufficient kinetic energy to inflict significant impact damage.


\section{Observations and Data Reduction} 
\label{sec:observations}

Photometry for A/2018 V3 was obtained using data from the telescopes shown in Table~\ref{tab:obs}; a complete list of the observing geometry is presented in Table~\ref{tab:data}.  All of the imaging data except for those from PS1 and CFHT Megacam were flattened in a standard manner using our image reduction pipeline.  Our photometric calibration accesses the Pan-STARRS \citep{magnier2013}, SDSS \citep{fukugita1996,doi2010}, and Gaia2 \citep{jordi2010} databases to provide a photometric zeropoint for each image using published color corrections to translate photometric bands.  The pipeline identifies image files by their instrument and generalizes access to their widely varying metadata.  The image reduction component bias-subtracts and flattens the data, applies the Terapix tool SCAMP \citep{bertin2006} to fit a precise World Coordinate System to the frame, and finally uses SExtractor \citep{bertin1996} to produce multi-aperture and automatic aperture target photometry.

\begin{deluxetable}{lccccc}[hbt!]
\scriptsize
\tablewidth{0pt}
\tablecaption{Observations\label{tab:obs}}
\tablecolumns{5}
\tablehead{
\colhead{Telescope/Instrument}
 & \colhead{Gain}
 & \colhead{RN}
 & \colhead{$''$/pix}
 & \colhead{Nts}
 & \colhead{\# $\dag$}
}
\startdata
Gemini/GMOS       & 2.27  & 3.32  & 0.161 & 4  & 23 \\
CFHT/MegaCam      & 1.634 & 3.00  & 0.185 & 11 & 32 \\
Pan-STARRS1/GPC1  & 1.26  & 7.46  & 0.260 & 4  & 11 \\
Pan-STARRS2/GPC2  & 1.0   & 7.46  & 0.257 & 9  & 38 \\
HCT/HFOSC 2       & 0.254 & 5.80  & 0.180 & 1  & 8  \\
ATLAS/STA-1600    & 2.0   & 11.0  & 1.86  & 26 & $^\ddag$\\
MPC               & ...   & ...   & ...   & 14 & $^\ddag$\\
\enddata
\tablecomments{$^{\dag}$Number of CCD images; $^{\ddag}$Received photometry (calibrated from ATLAS, uncalibrated from the MPC).}
\end{deluxetable}

\clearpage

\subsection{Gemini North 8m Telescope}
\label{sec:GN}

We obtained 23 images from the Gemini North telescope taken with the Gemini Multi-Object Spectrograph (GMOS) \citep{hook2004}, a mosaic of three 2048$\times$4176 Hamamatsu detectors, binned 2$\times$2.  The data were obtained through SDSS filters using queue service observing and were processed to remove the instrumental signature using DRAGONS, Gemini Observatory’s Python-based data reduction software \citep{labrie2019}.  Values in regions outside of the mosaic of detectors were manually converted to ``Not a Number'' (NaN) as necessary so that these areas were not counted as sky background.  For nights where the outer chips were non-photometric due to guide probe vignetting, we extracted only the central chip of the mosaic so that the zeropoint calibration would not be affected.

Data were obtained over four days: 2019 September 22; 2020 January 23 and 25; and 2020 July 22. We performed manual aperture photometry for each image by first utilizing a curve of growth to capture 99.5\% of the light for field stars with a similar brightness to the object.  We then measured the object through a range of apertures at 1-pixel step sizes and corrected the photometry to a 5$\arcsec$-radius aperture using the curve of growth.  This minimized the sky-dominated background noise for the object in each image.

\subsection{Canada-France-Hawai`i Telescope (CFHT)}
\label{sec:CFHT}

We obtained an additional 54 images using the CFHT MegaCam wide-field imager, an array of forty 2048$\times$4612 pixel CCDs with a plate scale of $0\farcs185$ per pixel and a 1.1 square degree FOV.  These images were obtained through queue service observing and covered 14 days spanning 2018 December through 2020 July.  Images were taken through the SDSS $r'$ filter ($\lambda_{\rm eff} = 0.640~\micron$, $\Delta\lambda = 0.148~\micron$) and CFHT's wideband $gri$ filter ($\lambda_{\rm eff} = 0.611~\micron$, $\Delta\lambda = 0.421~\micron$).  The data were pre-processed through CFHT's Elixir pipeline \citep{magnier2004} to remove the instrumental signature and then further processed through the pipeline described at the beginning of Section~\ref{sec:observations} for astrometric and photometric calibrations.

We manually inspected each image for issues.  Of the 54 images we received, 22 were not usable since the object fell too close to other sources in the frame.  The 32 images analyzed here consisted of 18 taken through the $r'$-band filter and 14 with the $gri$ filter.

\subsection{Pan-STARRS1 (PS1) and 2 (PS2)}
\label{sec:PS}
We received images and photometry data from Pan-STARRS1 (2019 January through 2020 April) and Pan-STARRS2 (2018 November to 2020 May).  Images were taken through the Pan-STARRS broadband $w$ filter ($\lambda_{\rm eff} = 0.608~\micron$, $\Delta\lambda = 0.382~\micron$), covering their $g$, $r$, and $i$ bandpasses.  All Pan-STARRS data were reduced by the Pan-STARRS Image Processing Pipeline (IPP; \citealt{magnier2020_IPP}) and calibrated against the Pan-STARRS database \citep{flewelling2020}.

\subsection{Asteroid Terrestrial-impact Last Alert System (ATLAS)}
\label{sec:ATLAS}
ATLAS provided us with 82 data points of photometry for A/2018 V3.  The data covered 2019 June through 2019 September and were taken through their broadband cyan ($\lambda_{{\rm eff},c} = 0.533~\micron$, $\Delta\lambda = 0.290~\micron$) and orange ($\lambda_{{\rm eff},o} = 0.679~\micron$, $\Delta\lambda = 0.261~\micron$) filters.  For nights where multiple data points were taken within a short observing interval ($\lesssim$30 minutes), we used a weighted average of the recorded magnitudes due to low signal-to-noise and large scatter.  These ATLAS data, averaged by date, are listed in Table~\ref{tab:data}.

\subsection{Himalayan Chandra Telescope}
\label{sec:HCT}
We obtained eight images taken on 2020 February 26 from the 2.01 m Himalayan Chandra Telescope (HCT) at Mt. Saraswati, Hanle, India.  The images were taken with the Himalaya Faint Object Spectrograph and Camera (HFOSC) and the new 4k$\times$4k e2V CCD with the Bessell/Cousins filter system.  These were then processed through the pipeline described at the beginning of Section~\ref{sec:observations}.  We performed manual aperture photometry for each image and used the weighted average of all data points taken for the night.

\subsection{NEOWISE}
\label{sec:NEOWISE}

We searched the Canadian Astronomy Data Centre's \citep{gwyn2012} archive using their Solar System Object Image Search (SSOIS) tool to search for data from the NEOWISE survey \citep{mainzer2014}. A total of 150 images of A/2018 V3 were found during four visits. The first visit, 2010 March 21-24, was during the cryogenic mission and the W1-W4 bands were available.  Without any cryogens after the first visit, only the two short-wavelength channels at $3.4~\micron$ (W1) and $4.6~\micron$ (W2) were available for the subsequent visits.

During visits on 2010 October 5-6, 2014 March 18-19, and 2014 October 1-2, A/2018 V3 was at heliocentric distances, $r = 22.43$, $15.98$, and $14.85$ au, respectively.  Because of the large distance and the inability of WISE to detect small objects at these distances, determining a meaningful upper limit to the size would not be possible from these observations and the data were not reduced.  However, there were two additional data points taken at much smaller heliocentric distances pre-perihelion, on 2019 January 2 ($r=3.49$ au, $\Delta=3.00$ au, $\rm{TA}=-103.83\degr$), and on 2019 July 20 ($r=1.53$ au, $\Delta=1.16$ au, $\rm{TA}=-41.46\degr$). 

\subsection{Other Data Sources}
\label{sec:other_sources}
We also gathered photometry data for A/2018 V3 from the Minor Planet Center's (MPC) database of observations.  R-band data came from nine different observatories and covered 14 days spanning 2018 November to 2020 April.  We averaged the reported magnitudes by observatory and by date and have assumed an average error on each measurement of $\pm0.25$.  These are included in the heliocentric light curve (see Figure~\ref{fig:lightcurve}) and the table of observational geometry (see Table~\ref{tab:data}).  However, the data only included the observed magnitude and filter, with no information on error, filter system, or photometry aperture.  For this reason, the MPC data were used only in constructing the heliocentric light curve, and not in the calculations and analysis.


\section{Analysis and Results}
\label{sec:analysis}

\subsection{Classification and Taxonomy}
\label{sec:taxonomy}

\subsubsection{Nucleus Colors}
\label{sec:colors}
Comparing the color of A/2018 V3 with those of other comets and asteroids allowed us to infer some of its surface properties.  Colors were calculated using the $5\arcsec$-radius aperture magnitudes from the 2020 January 23 and 25 Gemini data taken through SDSS $g'$, $r'$, $i'$, and $z'$ filters.  These are shown in Table~\ref{tab:colors}.  The $(g'-r')$ value from 2020 January 25 was then used to transform the data taken using other filter systems.  Magnitudes converted to SDSS $r'$ from other filters are denoted as $\bar{r}'$.  

To convert Pan-STARRS $w_{P1}$ filter magnitudes to $\bar{r}'$, we used the transformation given by \citet{tonry2012}:

\begin{equation}
\label{eq:PSw}
\bar{r}'= w_{p1} - 0.012 - 0.039(g'-r')
\end{equation}
\begin{equation}
\label{PSw_err}
    \sigma_{\bar{r}'}^2 = \sigma_{w_{p1}}^2 + (-0.039\sigma_{(g'-r')})^2 + 0.025^2
\end{equation}

Transformations from the ATLAS filter system magnitudes to SDSS $r'$ are found in \citet{tonry2018}.  Equations \ref{eq:ATLASc} and \ref{eq:ATLASc_err} refer to the ATLAS $c$-band values, while Equations \ref{eq:ATLASo} and \ref{eq:ATLASo_err} deal with the $o$-band magnitudes.  Again, the $(g'-r')$ value calculated from the 2020 January 25 Gemini data was used in the transformation.
\begin{equation}
\label{eq:ATLASc}
\bar{r}' = c - 0.47(g'-r')
\end{equation}
\begin{equation}
\label{eq:ATLASc_err}
\sigma_{\bar{r}'}^2 = \sigma_{c}^2 + (0.47\sigma_{(g'-r')})^2 + 0.01^2
\end{equation}
\begin{equation}
\label{eq:ATLASo}
\bar{r}' = o + 0.26(g'-r')
\end{equation}
\begin{equation}
\label{eq:ATLASo_err}
\sigma_{\bar{r}'}^2 = \sigma_{o}^2 + (0.26\sigma_{(g'-r')})^2 + 0.01^2
\end{equation}
To convert the CFHT $gri$ magnitudes to $\bar{r}'$, we used the following transformation from \citet{CADC_megapipe} along with the $(g'-i')$ color calculated from the 2020 January 25 Gemini data.
\begin{equation}
\label{eq:CFH-gri}
\bar{r}' = gri + 0.0068 - 0.2240(g'-i') + 0.0563(g'-i')^2
\end{equation}

\label{eq:CFH-gri_err}
\begin{multline}
    \sigma_{\bar{r}'}^2 = \sigma_{gri}^2 + (0.2240 \sigma_{(g'-i')})^2 +\\
    (0.1126(g'-i') \sigma_{(g'-i')})^2
\end{multline}
Finally, we used the following transformation given by \citet{lupton2005} to convert the Bessell/Cousins magnitudes to $\bar{r}'$.

\begin{equation}
\label{eq:bessell}
\bar{r}' = R + (0.1837(g'-r')) + 0.0971
\end{equation}
\textbf{\begin{equation}
\label{eq:bessell_err}
\sigma_{\bar{r}'}^2 = \sigma_{R}^2 + (0.1837\sigma_{(g'-r')})^2 + 0.0106^2
\end{equation}}

\begin{deluxetable}{ccc}[ht!]
\scriptsize
\tablewidth{0pt}
\tablecaption{A/2018 V3 colors\label{tab:colors}}
\tablecolumns{2}
\tablehead{
\colhead{}
 & \colhead{2020 Jan 23}
 & \colhead{2020 Jan 25}
}
\startdata
($g'-r'$)  & $0.649 \pm 0.031$  & $0.677 \pm 0.019$  \\
($g'-i'$)  & $0.965 \pm 0.031$  & $0.900 \pm 0.022$  \\
($r'-i'$)  & $0.316 \pm 0.028$  & $0.223 \pm 0.022$  \\
($r'-z'$)  & $0.506 \pm 0.025$  & $0.413 \pm 0.019$  \\
($i'-z'$)  & $0.190 \pm 0.026$  & $0.190 \pm 0.023$  \\
\enddata
\end{deluxetable}

\vspace{-1.0cm}
\subsubsection{Spectral Reflectivity}
\label{sec:refl}
We computed the spectral reflectivity, $R_{\lambda}$, at each wavelength normalized to the $g'$ filter using the color values calculated above along with the following equations

\begin{equation}
\label{eq:reflectivity}
R_{\lambda} = \frac{10^{-0.4(m_{\lambda}-M_{\lambda \odot})}}
	{10^{-0.4( m_{\rm ref} - M_{{\rm ref} \odot}) }}
\end{equation}
\begin{equation}
\label{eq:reflectivity_err}
	\sigma_{R{\lambda}} = 0.921 R_{\lambda} 
	\left[\sigma_{\lambda}^2 + \sigma_{\rm ref}^2 + \sigma_{\lambda \odot}^2 + \sigma_{{\rm ref} \odot}^2 \right]
\end{equation}

\noindent
Here, $m_{\lambda}$ is the object's magnitude in a specific filter with uncertainty $\sigma_\lambda$.  $M_{\lambda \odot}$ is the Sun's absolute magnitude for the same filter  with $\sigma_{\lambda \odot}$ uncertainty.  In the reference bandpass, $m_{\rm ref}$ is the object's magnitude and $\sigma_{\rm ref}$ its uncertainty, while $M_{{\rm ref} \odot}$ and $\sigma_{{\rm ref} \odot}$ are the solar magnitude and error in the reference filter.  

\begin{figure}[hbt]
\centerline{
\includegraphics[width=8.8cm]{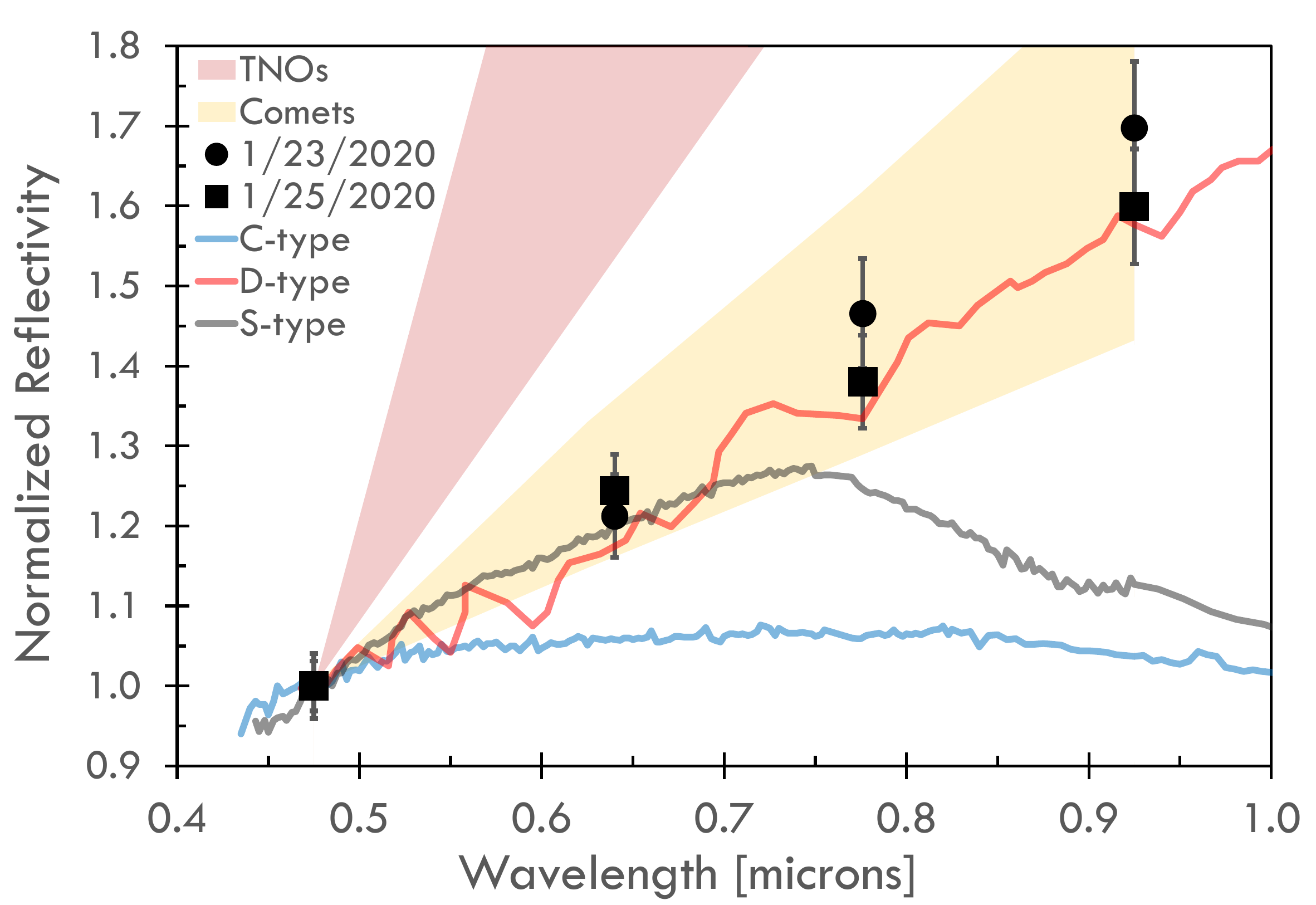}
}
\caption{The spectral reflectivity as calculated from Gemini data is consistent with the red color of comets and the D spectral type.  Data normalized to 1.0 at $\lambda=0.475~\micron$.}
\label{fig:reflectivity}
\end{figure}

For the solar magnitudes in the SDSS filters we used $g_{\odot}' = 5.12\pm0.020$, $r_{\odot}' = 4.68\pm0.028$, $i_{\odot}' = 4.57\pm0.035$, and $z_{\odot}' = 4.54\pm0.040$\footnote{\url{http://www.sdss.org/dr12/algorithms/ugrizvegasun/}}.  We then used the $(g'-r')$ color index calculated from the 2020 January 25 Gemini North data to normalize the spectral reflectivities to $\lambda=0.475~\micron$.  Spectral reflectivity values across the two days for which we had $g'r'i'z'$ magnitudes are shown in Figure~\ref{fig:reflectivity} and are consistent with the organic-rich red surface reflectivities of comets \citep{li2013,kelley2017}.

The normalized reflectivity slope $S'$ for this object was calculated from the averaged reflectivity values for the two days of Gemini data.  We used the method from \citet{jewitt1986}

\begin{equation}
\label{eq:spec-grad}
    S'\left[\frac{\%}{100 \rm nm}\right]=\left(\frac{20}{\Delta\lambda}\right) \frac{10^{0.4\Delta m}-1}{10^{0.4\Delta m}+1}
\end{equation}
\begin{equation}
\label{eq:spec-grad-err}
    \sigma_{S'}=\sigma_{\Delta m}\left[\frac{36.841\cdot10^{0.4 \Delta m}}{\Delta\lambda\left(10^{0.4\Delta m}+1\right)^2}\right]
\end{equation}
\noindent
where $\Delta m$ represents the difference between the object's color and the Sun's color for the bandpass $\Delta \lambda$, to calculate a gradient of $S'=10.6\pm0.9$ \%/100~nm.  This agrees with previously-calculated $S'$ values for asteroidal D spectral types \citep{hartmann1987,fitzsimmons1994,hicks2000,bus2002}.

\subsection{Searching for Activity}
\label{sec:activity}

\subsubsection{Surface Brightness Analysis Dust Production Limits}
\label{sec:dust}

We compared stellar surface brightness profiles with that of A/2018 V3 in deep stacked images to estimate upper limits on any dust production that could be produced by undetected ice-sublimation \citep{meech1996}.  The analysis was based on the heliocentric and geocentric locations of the object and assumed an average cometary dust grain size of 1~$\micron$ \citep{richter1995,horz2006,bentley2016,levasseurregourd2018}.

\begin{figure}[hbt!]
\centerline{
\includegraphics[width=8.5cm]{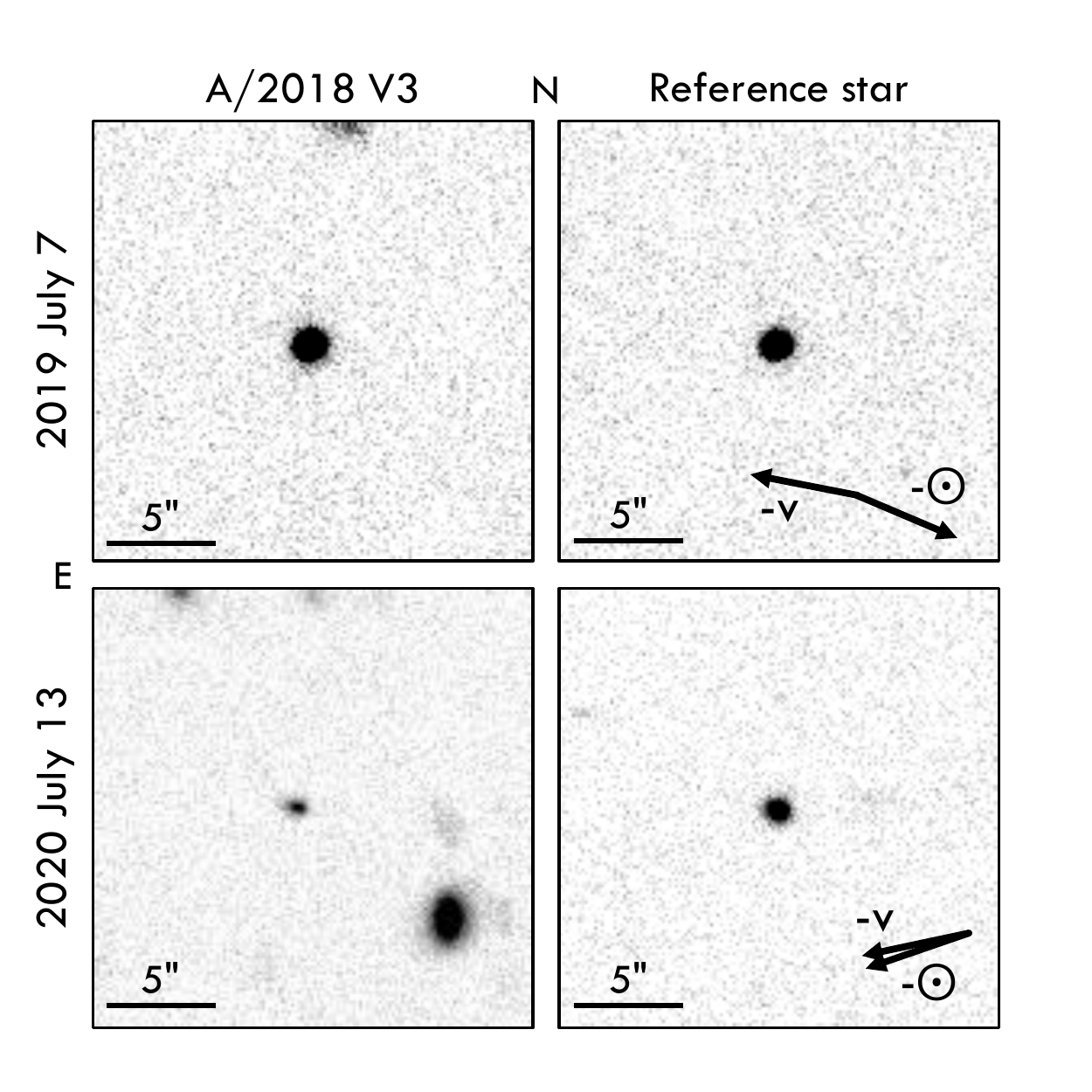}
}
\caption{Composite stacked images using CFHT frames; each image is $20\arcsec \times 20\arcsec$. $5\arcsec\approx5700$ km at a geocentric distance of 1.58 au in 2019; but $5\arcsec\approx16,000$ km at a geocentric distance of 4.32 au in 2020.  The image of A/2018 V3 is elongated E-W on 2020 July 13, roughly in the direction that a tail would be expected.
}
\label{fig:star-vs-obj}
\end{figure}

We used the IRAF \citep{tody1986} software to stack 4 frames taken 2019 July 7 with CFHT's MegaCam.  The frames were taken through $r'$-band filters with exposure times of 90 seconds each.  A/2018 V3 was at a heliocentric distance of 1.62 au and moving inbound toward the Sun.  The resulting image, seen in Figure~\ref{fig:star-vs-obj}, shows the object as a point source with no visibly-discernible activity, with a star from the same frame displayed alongside as reference.

\begin{figure}[hbt!]
\centerline{
\includegraphics[width=8.5cm]{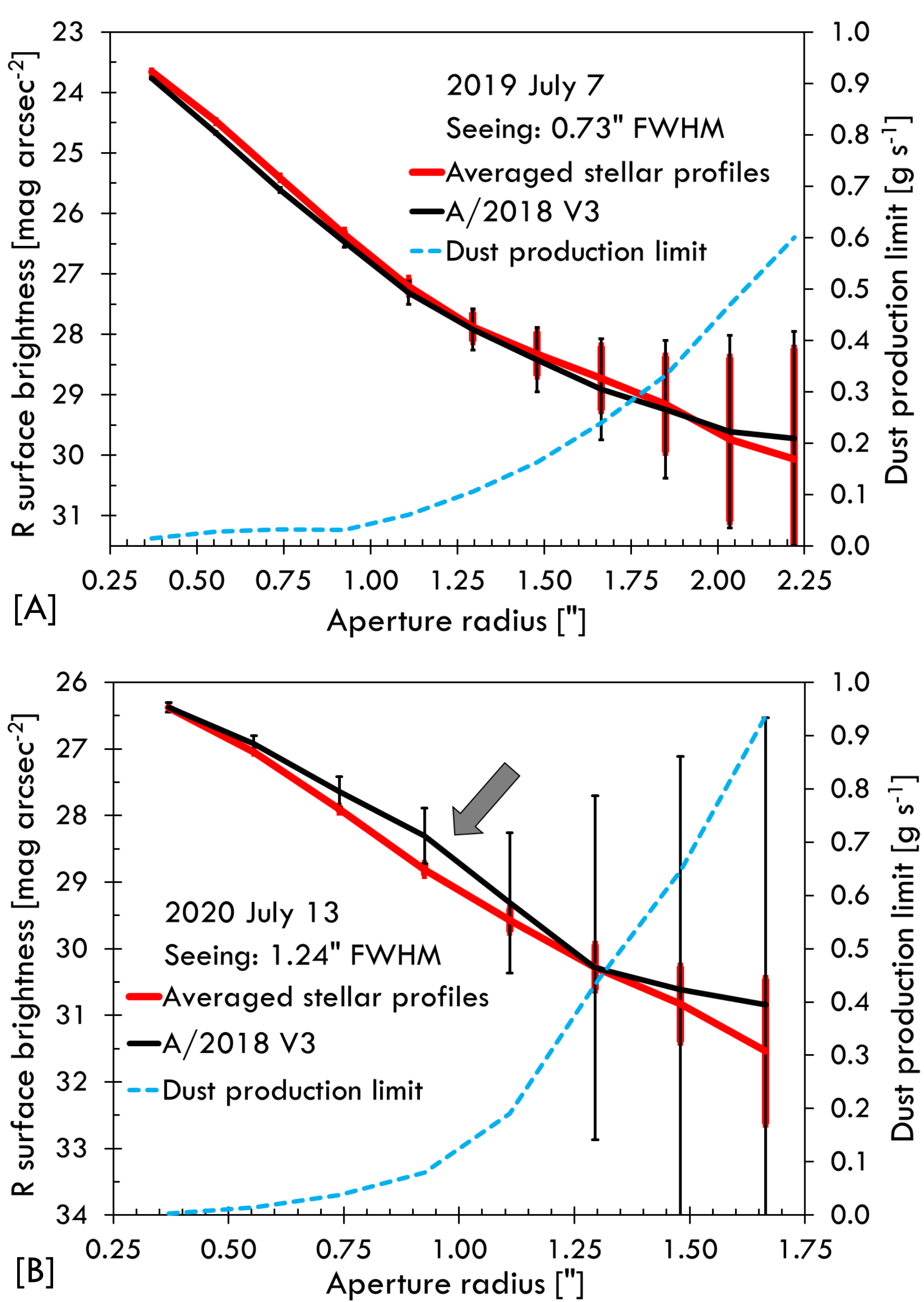}
}
\caption{[A] A/2018 V3's pre-perihelion ($\rm{TA}=-49.5\degr$) surface brightness profile is indistinguishable from that of stars, which is consistent with a point source showing no dust activity.  [B] Post-perihelion ($\rm{TA}=110.5\degr$), however, the surface brightness profile of A/2018 V3 is broader, consistent with the extension seen in Figure~\ref{fig:star-vs-obj}. The arrow indicates possible low-level dust production for the object near 0.9$\arcsec$.}
\label{fig:sb}
\end{figure}

We computed an azimuthally-averaged radial surface brightness profile for both the object and each of three field stars of comparable brightness.  The field star fluxes were normalized and averaged at each radial distance to form an average stellar profile for comparison with the profile of A/2018 V3.  The resulting surface brightness profiles for the object and for the average of the three field stars are shown in Figure~\ref{fig:sb}A.  No activity was apparent in either the composite stacked image or in a difference between the surface brightness profiles.  From this, we calculated an average upper limit for dust production of $\approx$0.13 g $\rm s^{-1}$ at a distance of 1.2$\arcsec$ from the object's core, using the method described in \citet{meech2003}.  The full curve for the calculated dust production limit is shown as the dashed line in Figure~\ref{fig:sb}A.

We created a second image stack of 3 frames from CFHT data taken 2020 July 13 using the $gri$ filter.  Individual frames showed a slight east-west elongation in the object, whereas stellar profiles in the same frame were round, as seen in the lower row of Figure~\ref{fig:star-vs-obj}.  The object was moving north to south across the telescope's field of view and as such, significant east-west trailing would have been unlikely.  Figure~\ref{fig:sb}B shows the surface brightness comparison for the 2020 July 13 data.  Here, the object's radial profile is compared against the averaged profile for nine field stars.

The small bump in the object's brightness profile (indicated by the arrow in Figure~\ref{fig:sb}B) at $0.9\arcsec$ likely corresponds to the small east-west elongation noted in the deep image stack.  However, since both brightness profiles fall within their respective margins of error, we cannot say with certainty that this indicates activity.

We calculated an upper limit on dust production of $\approx$0.68 g $\rm s^{-1}$ at a distance of $1.5\arcsec$ from the object's core, again using the method from \citet{meech2003}.  This is indicated by the dashed line in Figure~\ref{fig:sb}B.  Using this upper limit on dust production, we calculated an active fractional area of $7.17\times10^{-5}$ for a spherical nucleus with average radius $r\approx2$ km (discussed further in $\S$~\ref{sec:models}).  This would correspond to a maximum surface area of 0.0036 km$^2$ that could be active.

\begin{figure*}[htb!]
\centerline{
\includegraphics[width=18cm]{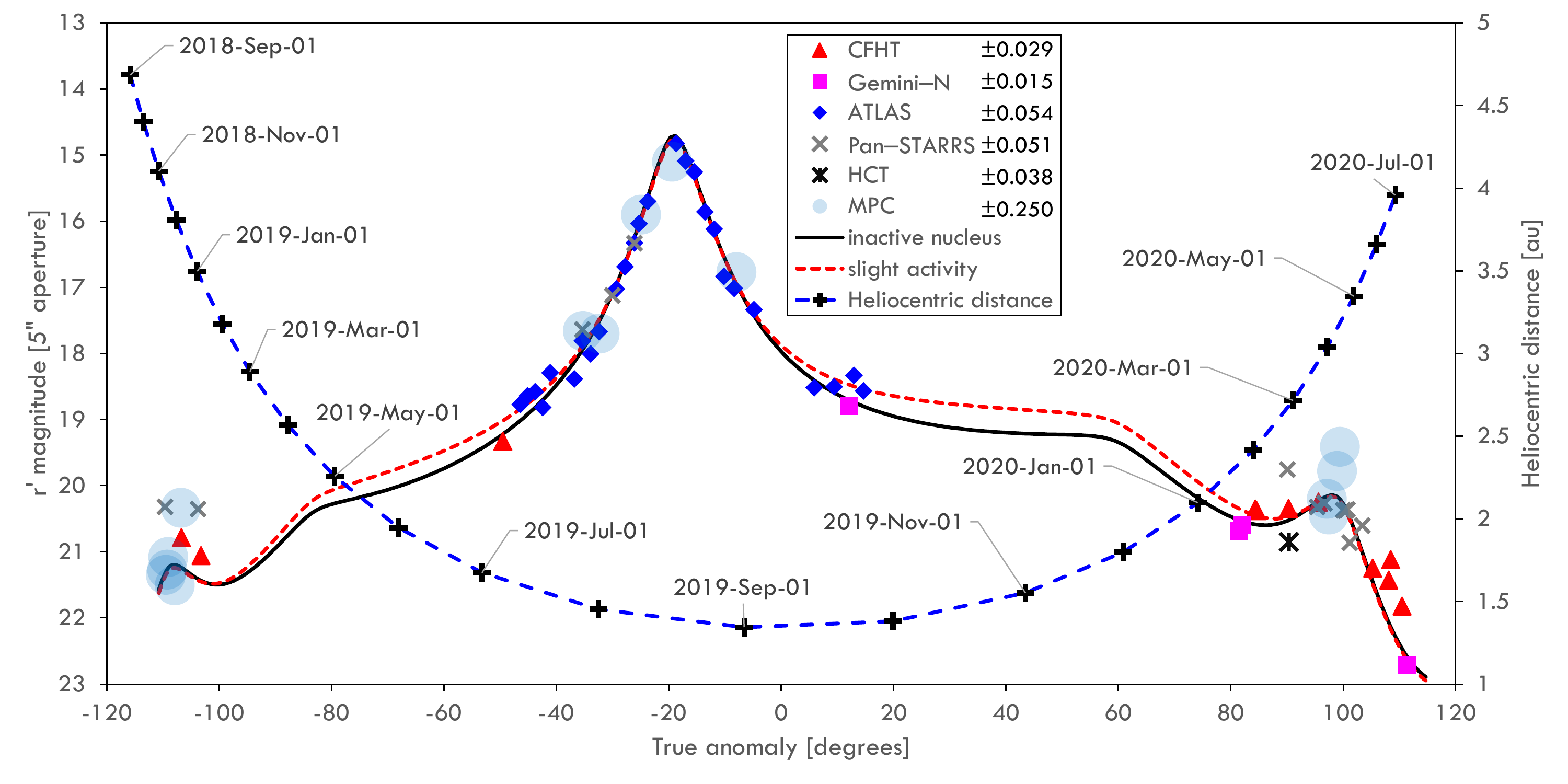}
}
\caption{This heliocentric light curve shows that our observations are consistent with only a very weakly active nucleus. The peak brightness near $\rm{TA}=-20\degr$ coincides with the object's closest geocentric distance.  The average errors for each data set are indicated in the legend.}
\label{fig:lightcurve}
\end{figure*}

If A/2018 V3 was indeed active and sublimating around 2020 July 13, this amount of activity was not sufficient to unambiguously lift optically significant dust grains.  The lack of significant deviation from either light curve in Figure~\ref{fig:lightcurve} (discussed in $\S$~\ref{sec:models}), along with the detailed image inspection, all point to either an inactive or a barely active nucleus.

\subsubsection{Sublimation Model}
\label{sec:models}

We constructed a heliocentric light curve (Figure~\ref{fig:lightcurve}) from the SDSS $r'$ magnitudes as a function of the object's true anomaly.  This served as the basis for understanding whether or not A/2018 V3 was active, particularly around perihelion and immediately following.  We also used the light curve to find any data points that deviated significantly from the trend.  We conducted a careful examination of the corresponding images to rule out contamination from background objects, cosmic rays, or bad pixels.  The curves fit to the data points were determined using the surface ice sublimation model as in \citet{meech1986}.

The sublimation model enabled us to explore the light curve and to search for and quantify any possible activity.  This model has been used successfully to explain the behavior of comets where we do not have a lot of detailed information, and has proven very successful in predicting and explaining the behavior of well-observed mission targets \citep{meech2011,snodgrass2013,meech2017}. The sublimation model computes the amount of gas sublimating from an icy surface exposed to solar heating, as described in detail in \citet{meech2017}.  This escaping gas flow can drag dust grains from the nucleus and thereby modify the observed brightness of the object.  By analyzing the shape of the heliocentric light curve, the model can distinguish between H$_2$O, CO, and CO$_2$ driven activity based on the total brightness within a fixed aperture.  This brightness includes radiation scattered from both the nucleus and any of the escaping dust grains.  Model free parameters include nucleus properties (radius, density, albedo, emissivity), dust properties (grain size, density, phase function), and the fractional active area.

The light curves in Figure~\ref{fig:lightcurve} show the results of two model runs based on the 2020 July 13 data: one assuming no activity (i.e. representing the bare nucleus), and the other consistent with the dust production determined from the surface brightness analysis (Section~\ref{sec:dust}).  Both models assumed an average cometary albedo of 0.04 \citep{li2013} an average dust grain size of 1~$\micron$ \citep{levasseurregourd2018}, and a dust to gas mass ratio of 1 \citep{marschall2020}.  By using the sublimation and dust models in tandem, we gradually adjusted the free parameters of each until the resulting model curves visually matched the data.  We then used least-squares fitting to calculate an average nucleus radius of $\approx$2.06 km for an inactive object, or $\approx$2 km for a nucleus with a fractional active area of $7.17\times 10^{-5}$.  The curve fit to this weakly-active nucleus implied a water production rate of $2.3\times 10^{22}$ molec/s.  

\subsection{NEOWISE Size Estimate}
\label{sec-WISE}

We also used the {\it NEOWISE} data to determine the effective radius of A/2018 V3.  As mentioned in Section~\ref{sec:NEOWISE}, images were available only in the W1 and W2 bands.  The WISE data have all been processed through the WISE science data pipeline \citep{wright2010} to bias-subtract and flatten the images, and to remove artifacts.  The images are then stacked using the comet's apparent rate of motion \citep{bauer2015}.  Aperture photometry is converted to fluxes using the {\it WISE} zeropoints and appropriate color temperature corrections \citep{wright2010}.  These corrections are temperature-dependent, and an initial temperature estimate is required based on the expected blackbody temperature for the heliocentric distance of the observation.\\

\begin{figure}[hbt!]
\centerline{
\includegraphics[width=8cm]{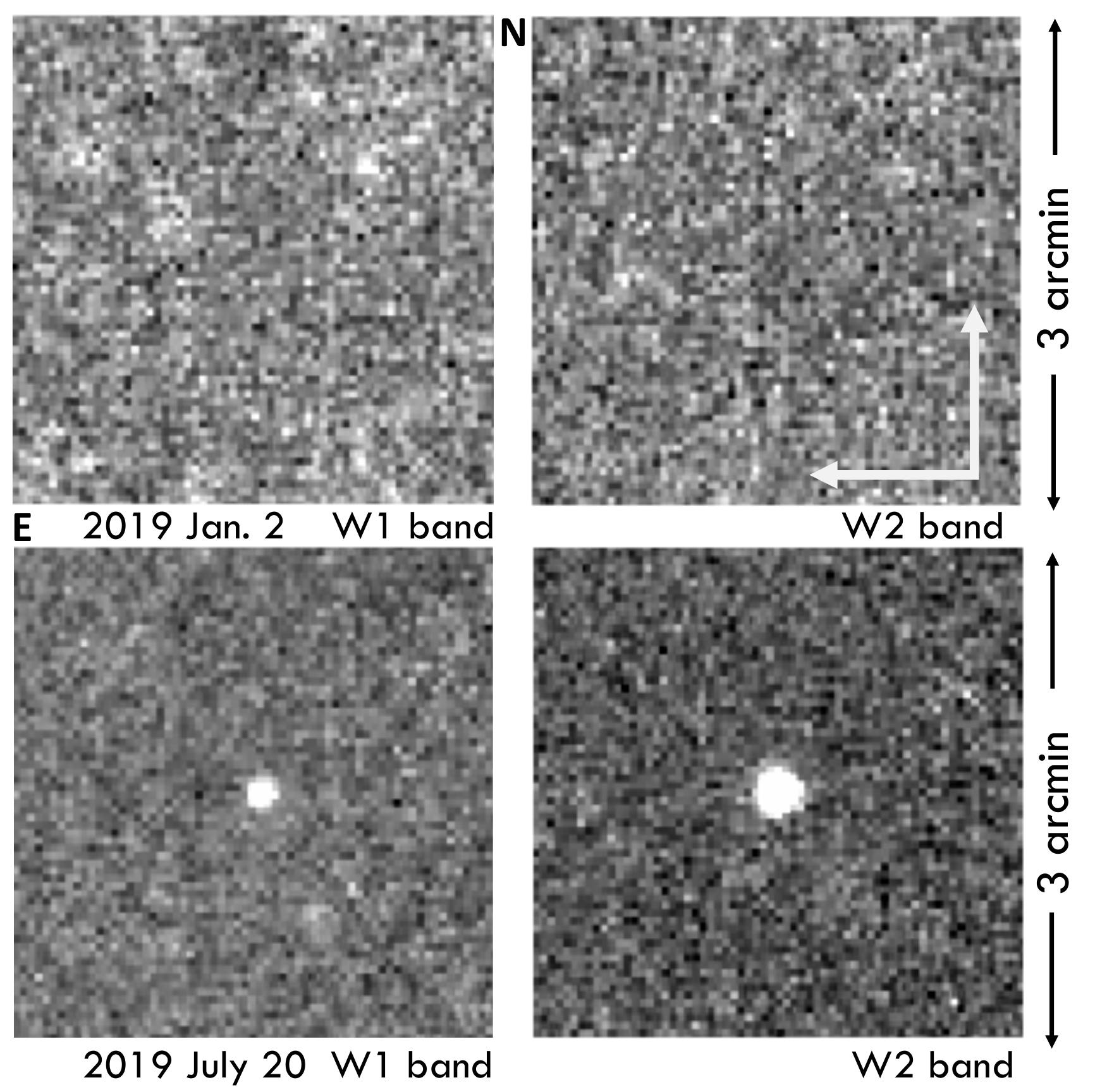}
}
\caption{The W1 and W2 co-added images from 2019 January 2 and July 20. There is no significant signal in either of the January frames, but the July frames clearly both show high signal-to-noise, with the W2 image showing the greatest. The images show no evidence of extended emission.}
\label{fig:NEOWISEim}
\end{figure}

\begin{figure}[hbt!]
\centerline{\includegraphics[width=8.0cm]{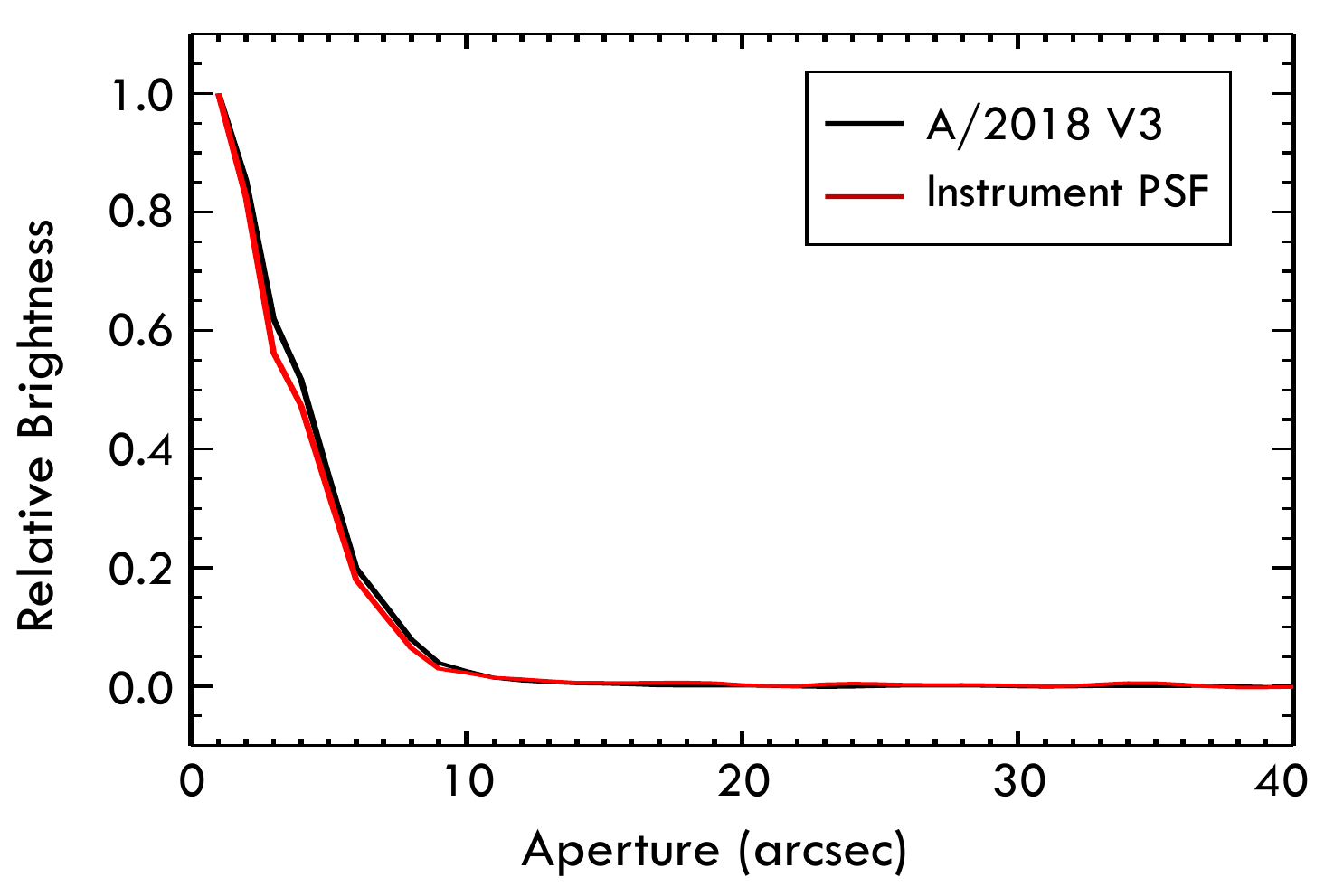}}
\caption{The NEOWISE W2 profile of A/2018 V3 relative to the instrumental profile in the same bandpass. Both profiles have been normalized to the peak counts. The profiles are nearly a perfect match, indicating no coma detection.}
\label{fig:NEOWISEsbp}
\end{figure}

The data from 2019 January 2 show no signal in either band down to a 1-$\sigma$ magnitude of 18.4, which yields a 3-$\sigma$ upper limit of 12 km for the nucleus radius (which is unconstraining).  However, detections were identified in data from 8 exposures taken on 2019 July 20, centered at 01:23:30.585 UT (Figure~\ref{fig:NEOWISEim}). A/2018 V3 was clearly not active at the time of these observations as shown by the profile comparison in Figure~\ref{fig:NEOWISEsbp}.  Using the NEATM thermal model \citep{harris1998,masiero2017,mainzer2019PDS} along with the observed fluxes, we derived a radius of $2.7 \pm 0.9$ km.  Coupled with the observations provided here, this yields a surface reflectivity of $\sim 0.03_{-0.01}^{+0.02}$, consistent with the derived value used in Section~\ref{sec:models}.

\section{Discussion}
\label{sec:discuss}
\begin{figure}[hbt]
\centerline{
\includegraphics[width=8.8cm]{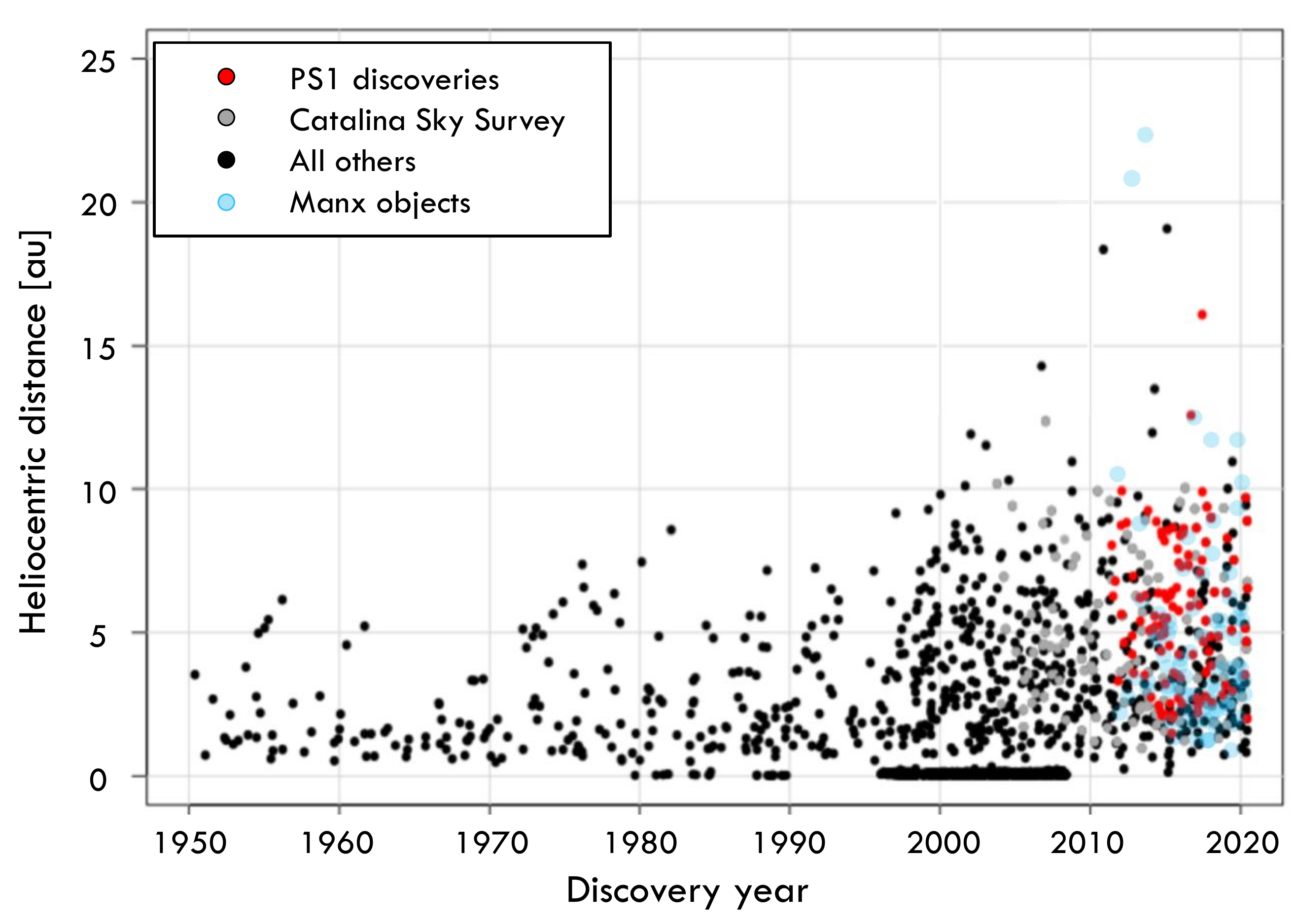}
}
\caption{The advent of large all-sky surveys in the mid-1990s increased both the discovery rate and the distance at which comets are discovered.  Because Manx comets are fainter and harder to detect than LPCs, most are discovered only within $\sim$4-5 au of the Sun.}
\label{fig:Manx-disc}
\end{figure}

Figure~\ref{fig:Manx-disc} shows how the large survey projects such as Pan-STARRS and the Catalina Sky Survey have dramatically increased the number of known small solar system bodies in the last 20 years.  These surveys have enabled detections of an increasing percentage of ever-smaller solar system bodies in Earth's neighborhood.  No longer is there such a heavy sampling bias towards large, bright objects; but this also means that there must be many more undetected, low-albedo objects in near-Earth space.

Both \citet{mainzer2011} and \citet{harris2015} have estimated that about 90\% of near-Earth asteroids (NEAs) larger than one kilometer in diameter have been discovered.  However, the latter study also noted the inherent bias against detecting NEAs with high eccentricities and long orbital periods.  Manxes and LPCs have just such orbital parameters; thus, long-period objects that also qualify as NEAs are likely underrepresented in asteroid surveys.

Such high-eccentricity orbits also imbue Manxes and LPCs with much higher velocities near perihelion than the usual near-Earth object.  At its closest geocentric approach (0.37 au), A/2018 V3 was moving at 64 km/s relative to Earth\footnote{JPL Small-Body Database Browser: \url{https://ssd.jpl.nasa.gov/sbdb.cgi}}; the average NEO velocity is approximately 21 km/s relative to Earth \citep{shannon2015}.  A/2018 V3 has passed both Earth and the Sun without incident and will not return to the inner solar system for another 1300 years, but given that Manxes are likely underrepresented in surveys, the threat from long-period objects may be higher than previously believed.

Figure~\ref{fig:Manx-disc} demonstrates that most of the Manx objects discovered thus far have only been detected within $\sim$4 au of the Sun.  A/2018 V3 was a typical Manx in this regard: at its discovery in 2018 November, it was almost exactly 4 au from perihelion.  This translated to about nine months between discovery and closest geocentric approach.  The inherent nature of these objects means that we could have very little time to implement mitigation measures should a new Manx be discovered on an impact trajectory with Earth \citep{napier2004,nuth2018}.

We realize that with present technology, it would be cost-prohibitive and probably impossible to catalog every single LPC and Manx.  Some have argued that the probability of an LPC impacting Earth is small enough to rate a lower priority for planetary defense than the asteroids \citep{shannon2015,stokes2017}.  We argue that this group of inconspicuous, yet potentially-dangerous objects has not yet been fully characterized and thus warrants further attention.  Additional information will lead to more accurate estimates of overall population size and give us a better idea of where to best allocate limited planetary defense resources.

The study of A/2018 V3 has demonstrated that we do not yet have a comprehensive picture of these objects.  A/2018 V3 had a red spectral slope characteristic of comets, yet even at 1.34 au from the Sun, well within the range where water ice sublimation would be expected for comets, did not display an obvious tail.  However, possible activity was seen as a small dust coma approximately 10 months after perihelion in July 2020.  Other Manx objects are weakly active while inbound \citep{meech2014}; still others appear to be rocky inner solar system material and yet display low-level activity consistent with water ice sublimation \citep{meech2016}.  It will be important to determine whether or not Manxes represent a distinct and diverse category of small solar system bodies that may have formed near the solar system ice-line and subsequently ejected to the Oort cloud during formation, or if they represent long period comets that have lost their surface volatiles.

If Manx objects do indeed represent a cohesive population, an in-depth study will provide insight as to where they fall in the context of solar system dynamics.  We can then compare the composition and distribution of Oort Cloud populations against what would be expected based on the various models of solar system dynamics.  These small, unassuming Manx objects can thus impose limits on the magnitude of giant planet migration early in the solar system's history, and thereby help constrain the current models.\\

\acknowledgments

CPYP acknowledges Dr. Dominic Piro for software expertise and assistance with numerical models, and also acknowledges support from the National Science Foundation (NSF) Research Experience for Undergraduate program at the Institute for Astronomy, University of Hawai`i at M\={a}noa, funded through NSF grant 6104374. KJM, JTK, and JVK acknowledge support through awards from the National Science Foundation AST1413736 and AST1617015, and from the National Aeronautics and Space Administration (NASA) under grant NASA-80NSSC18K0853. RJW acknowledges support by NASA under grant NNX14AM74G issued through the SSO Near Earth Object Observations Program. LD acknowledges support by NASA under grants NNX12AR55G and NNX14AM74G.

Data were acquired using the PS1 System operated by the PS1 Science Consortium (PS1SC) and its member institutions. The PS1 Surveys have been made possible by contributions from PS1SC member Institutions and NASA through Grant NNX08AR22G, the NSF under Grant No. AST-123886, the Univ. of MD, and Eotvos Lorand Univ.  This work also uses data from the ATLAS project, funded through NASA grants NN12AR55G, 80NSSC18K0284, and 80NSSC18K1575, with the IfA at the University of Hawai`i, and with  contributions from the Queen's University Belfast, STScI, and the South African Astronomical Observatory.

This publication also makes use of data products from the Near-Earth Object Wide-field Infrared Survey Explorer (NEOWISE), which is a joint project of the Jet Propulsion Laboratory/California Institute of Technology and the University of Arizona. NEOWISE is funded by the National Aeronautics and Space Administration.

Based also in part on observations obtained with MegaPrime/MegaCam, a joint project of CFHT and CEA/DAPNIA, at the Canada-France-Hawaii Telescope (CFHT) which is operated by the National Research Council (NRC) of Canada, the Institute National des Science de l'Univers of the Centre National de la Recherche Scientifique (CNRS) of France, and the University of Hawai`i.   This research used the facilities of the Canadian Astronomy Data Centre operated by the National Research Council of Canada with the support of the Canadian Space Agency.

We thank the staff of IAO, Hanle and CREST, Hosakote, that made these observations possible.  IAO and CREST are operated by the Indian Institute for Astrophysics, Bangalore.

This research has also made use of data and/or services provided by the International Astronomical Union's Minor Planet Center.

\startlongtable
\begin{deluxetable*}{lccccccccc}
\tablecaption{Observing Geometry and Photometry \label{tab:data}}
\tablehead{
\colhead{UT Date                    } &
\colhead{JD\tablenotemark{a}        } &
\colhead{$r$\tablenotemark{b}       } &
\colhead{$\Delta$\tablenotemark{b}  } &
\colhead{$\alpha$\tablenotemark{b}  } &
\colhead{TA\tablenotemark{c}        } &
\colhead{Filt                       } &
\colhead{\# Images                  } &
\colhead{mag$\pm$$\sigma$\tablenotemark{d}} &
\colhead{r$_{mag}$$'$$\pm$$\sigma$\tablenotemark{d}}
}
\startdata
\multicolumn{10}{l}{\bf Gemini North data} \\
\hline
2019/09/22 & 8748.72873 & 1.354  & 1.303  & 44.354 &  12.04  & r$'$     &  2 &    $18.796 \pm 0.014$ & \\
2020/01/23 & 8872.14957 & 2.324  & 2.686  & 21.098 &  81.50  & r$'$     &  1 &    $20.691 \pm 0.019$ & \\
2020/01/23 & 8872.15054 & 2.324  & 2.686  & 21.098 &  81.50  & i$'$     &  1 &    $20.375 \pm 0.020$ & \\
2020/01/23 & 8872.15226 & 2.324  & 2.686  & 21.098 &  81.50  & z$'$     &  2 &    $20.185 \pm 0.016$ & \\
2020/01/23 & 8872.15449 & 2.324  & 2.686  & 21.098 &  81.50  & g$'$     &  1 &    $21.340 \pm 0.024$ & \\
2020/01/25 & 8874.13988 & 2.345  & 2.668  & 21.418 &  82.09  & i$'$     &  1 &    $20.374 \pm 0.018$ & \\
2020/01/25 & 8874.14160 & 2.345  & 2.668  & 21.419 &  82.09  & z$'$     &  2 &    $20.184 \pm 0.014$ & \\
2020/01/25 & 8874.14460 & 2.345  & 2.668  & 21.419 &  82.09  & g$'$     &  3 &    $21.274 \pm 0.013$ & \\
2020/01/25 & 8874.14691 & 2.345  & 2.668  & 21.419 &  82.09  & Y$'$     &  2 &    $20.246 \pm 0.029$ & \\
2020/01/25 & 8874.15129 & 2.345  & 2.668  & 21.420 &  82.09  & r$'$     &  3 &    $20.597 \pm 0.013$ & \\
2020/07/22 & 9052.77292 & 4.162  & 4.550  & 12.393 &  111.34 & r$'$     &  5 &    $22.711 \pm 0.014$ & \\
\hline
\multicolumn{10}{l}{\bf CFHT data} \\
\hline
2018/12/09 & 8461.90062 & 3.727  & 2.898  & 9.302  &  -106.75 & r$'$    &  3 &    $20.783 \pm 0.029$ & \\
2019/01/06 & 8489.78282 & 3.446  & 3.039  & 15.902 &  -103.26 & r$'$    &  3 &    $21.057 \pm 0.038$ & \\
2019/07/07 & 8672.10057 & 1.623  & 1.579  & 36.992 &  -49.54  & r$'$    &  4 &    $19.327 \pm 0.009$ & \\
2020/02/02 & 8882.15529 & 2.428  & 2.588  & 22.363 &   84.36  & r$'$    &  2 &    $20.541 \pm 0.021$ & \\
2020/02/03 & 8883.16645 & 2.438  & 2.577  & 22.440 &   84.63  & r$'$    &  2 &    $20.835 \pm 0.018$ & \\
2020/02/26 & 8906.15469 & 2.678  & 2.323  & 21.357 &   90.30  & gri     &  2 &    $20.581 \pm 0.015$ & $20.34 \pm 0.02$ \\
2020/03/23 & 8932.10234 & 2.948  & 2.140  & 13.373 &   95.59  & r$'$    &  2 &    $20.247 \pm 0.016$ & \\
2020/05/26 & 8995.80694 & 3.599  & 3.010  & 14.389 &   105.23 & r$'$    &  2 &    $21.243 \pm 0.040$ & \\
2020/06/20 & 9020.75373 & 3.848  & 3.687  & 15.310 &   108.13 & gri     &  4 &    $21.665 \pm 0.034$ & $21.42 \pm 0.03$ \\
2020/06/23 & 9023.77276 & 3.878  & 3.771  & 15.191 &   108.45 & gri     &  5 &    $21.354 \pm 0.030$ & $21.11 \pm 0.03$ \\
2020/07/13 & 9043.79350 & 4.075  & 4.317  & 13.522 &   110.49 & gri     &  3 &    $22.062 \pm 0.055$ & $21.82 \pm 0.05$ \\
\hline
\multicolumn{10}{l}{\bf Pan-STARRS1 data}\\
\hline
2019/01/02 & 8485.80048 & 3.486  & 3.007  & 15.237 & -103.79  & w$_{p1}$ &  1 &    $20.374 \pm 0.086$  &  $20.34 \pm 0.09$ \\ 
2019/08/04 & 8700.09003 & 1.435  & 0.649  & 38.814 & -30.04   & w$_{p1}$ &  2 &    $17.160 \pm 0.014$  &  $17.12 \pm 0.02$ \\ 
2020/04/18 & 8958.06430 & 3.216  & 2.261  & 6.651  &  99.99   & w$_{p1}$ &  4 &    $20.409 \pm 0.046$  &  $20.37 \pm 0.05$ \\
2020/04/23 & 8962.85430 & 3.265  & 2.324  & 7.383  &  100.73  & w$_{p1}$ &  3 &    $20.396 \pm 0.087$  &  $20.36 \pm 0.09$ \\
\hline
\multicolumn{10}{l}{\bf Pan-STARRS2 data}\\
\hline
2018/11/12 & 8434.91595 & 3.994  & 3.051  & 4.963  & -109.66 & w$_{p2}$ &  4 &    $20.356 \pm 0.052$   & $20.32 \pm 0.05$  \\ 
2019/07/28 & 8693.09891 & 1.475  & 0.875  & 42.279 & -35.36  & w$_{p2}$ &  7 &    $17.675 \pm 0.004$   & $17.64 \pm 0.01$  \\ 
2019/08/09 & 8705.06070 & 1.411  & 0.508  & 31.594 & -26.09  & w$_{p2}$ &  3 &    $16.369 \pm 0.006$   & $16.33 \pm 0.02$  \\ 
2020/02/25 & 8905.13669 & 2.667  & 2.334  & 21.529 &  90.07  & w$_{p2}$ &  3 &    $19.793 \pm 0.051$   & $19.76 \pm 0.05$  \\ 
2020/03/22 & 8931.07134 & 2.937  & 2.143  & 13.789 &  95.40  & w$_{p2}$ &  3 &    $20.359 \pm 0.063$   & $20.32 \pm 0.06$  \\ 
2020/03/29 & 8938.01692 & 3.009  & 2.136  & 10.972 &  96.66  & w$_{p2}$ &  2 &    $20.302 \pm 0.067$   & $20.26 \pm 0.07$  \\
2020/04/21 & 8960.86413 & 3.244  & 2.296  & 7.013  &  100.42 & w$_{p2}$ &  8 &    $20.405 \pm 0.034$   & $20.37 \pm 0.04$  \\
2020/04/26 & 8965.82409 & 3.295  & 2.368  & 8.050  &  101.17 & w$_{p2}$ &  4 &    $20.899 \pm 0.060$   & $20.86 \pm 0.06$  \\
2020/05/12 & 8981.79239 & 3.458  & 2.674 & 12.027  &  103.42 & w$_{p2}$ &  4 &    $20.641 \pm 0.055$   & $20.60 \pm 0.06$  \\
\hline
\multicolumn{10}{l}{\bf HCT data} \\
\hline
2020/02/26 & 8906.40805 & 2.680  & 2.321 & 21.314  &  90.35   & R$_{B}$ &  8 &    $20.631 \pm 0.037$  &  $20.85 \pm 0.04$ \\
\hline
\multicolumn{10}{l}{\bf ATLAS data} \\
\hline
2019/07/12 & 8677.02859 & 1.585  & 1.416  & 39.090 & -46.46  & o	   &  2 &    $18.598 \pm 0.093$  & $18.77 \pm 0.09$ \\
2019/07/14 & 8679.04963 & 1.570  & 1.349  & 39.867 & -45.15  & o	   &  2 &    $18.470 \pm 0.112$  & $18.64 \pm 0.11$ \\
2019/07/16 & 8681.10287 & 1.555  & 1.280  & 40.588 & -43.79  & o	   &  3 &    $18.406 \pm 0.088$  & $18.58 \pm 0.09$ \\
2019/07/18 & 8683.10049 & 1.540  & 1.213  & 41.209 & -42.45  & o	   &  1 &    $18.645 \pm 0.161$  & $18.82 \pm 0.16$ \\
2019/07/20 & 8685.04042 & 1.527  & 1.147  & 41.718 & -41.12  & o       &  1 &    $18.119 \pm 0.144$  & $18.29 \pm 0.14$ \\
2019/07/26 & 8691.09842 & 1.487  & 0.942  & 42.449 & -36.83  & o	   &  4 &    $18.210 \pm 0.033$  & $18.38 \pm 0.03$ \\
2019/07/28 & 8693.05001 & 1.475  & 0.877  & 42.287 & -35.40  & o	   &  4 &    $17.636 \pm 0.033$  & $17.81 \pm 0.03$ \\
2019/07/30 & 8695.01820 & 1.463  & 0.811  & 41.841 & -33.93  & c	   &  4 &    $18.327 \pm 0.034$  & $18.01 \pm 0.03$ \\
2019/08/01 & 8697.03012 & 1.452  & 0.746  & 41.016 & -32.41  & o	   &  4 &    $17.498 \pm 0.031$  & $17.67 \pm 0.03$ \\
2019/08/05 & 8700.98035 & 1.431  & 0.622  & 37.901 & -29.35  & o	   &  4 &    $16.851 \pm 0.015$  & $17.03 \pm 0.02$ \\
2019/08/07 & 8702.95738 & 1.421  & 0.565  & 35.325 & -27.78  & c	   &  4 &    $17.013 \pm 0.015$  & $16.69 \pm 0.02$ \\
2019/08/09 & 8705.00221 & 1.411  & 0.510  & 31.713 & -26.14  & o	   &  5 &    $16.156 \pm 0.010$  & $16.33 \pm 0.01$ \\
2019/08/10 & 8706.01279 & 1.407  & 0.484  & 29.519 & -25.32  & c	   &  4 &    $16.356 \pm 0.010$  & $16.04 \pm 0.01$ \\
2019/08/12 & 8707.93176 & 1.398  & 0.442  & 24.560 & -23.75  & o	   &  4 &    $15.529 \pm 0.010$  & $15.70 \pm 0.01$ \\
2019/08/18 & 8713.97130 & 1.376  & 0.374  & 11.501 & -18.70  & o	   &  3 &    $14.651 \pm 0.007$  & $14.82 \pm 0.01$ \\
2019/08/20 & 8715.92015 & 1.369  & 0.379  & 16.681 & -17.03  & o	   &  4 &    $14.915 \pm 0.006$  & $15.09 \pm 0.01$ \\
2019/08/22 & 8717.73993 & 1.364  & 0.397  & 23.275 & -15.47  & o	   &  3 &    $15.083 \pm 0.010$  & $15.26 \pm 0.01$ \\
2019/08/24 & 8719.90087 & 1.358  & 0.431  & 30.596 & -13.59  & o	   &  3 &    $15.685 \pm 0.009$  & $15.86 \pm 0.01$ \\
2019/08/26 & 8721.78821 & 1.354  & 0.471  & 35.859 & -11.94  & c	   &  3 &    $16.445 \pm 0.014$  & $16.12 \pm 0.02$ \\
2019/08/28 & 8723.75967 & 1.350  & 0.519  & 40.147 & -10.21  & o	   &  1 &    $16.660 \pm 0.050$  & $16.83 \pm 0.05$ \\
2019/08/30 & 8725.81659 & 1.347  & 0.575  & 43.452 & -8.39   & c	   &  1 &    $17.335 \pm 0.110$  & $17.01 \pm 0.11$ \\
2019/09/03 & 8729.77504 & 1.342  & 0.693  & 47.224 & -4.86   & c	   &  4 &    $17.657 \pm 0.028$  & $17.34 \pm 0.03$ \\
2019/09/15 & 8741.76290 & 1.343  & 1.079  & 47.557 &  5.87   & o	   &  2 &    $18.343 \pm 0.094$  & $18.52 \pm 0.09$ \\
2019/09/19 & 8745.74047 & 1.349  & 1.207  & 45.893 &  9.40   & o	   &  7 &    $18.331 \pm 0.050$  & $18.50 \pm 0.05$ \\
2019/09/23 & 8749.75147 & 1.357  & 1.335  & 43.784 &  12.93  & c	   &  4 &    $18.654 \pm 0.066$  & $18.33 \pm 0.07$ \\
2019/09/25 & 8751.71820 & 1.361  & 1.397  & 42.637 &  14.64  & o	   &  1 &    $18.393 \pm 0.142$  & $18.57 \pm 0.14$ \\
\hline
\multicolumn{10}{l}{\bf MPC data$^{\dag}$}\\
\hline
2018/11/14 & 8436.8     & 3.975  & 3.030  & 4.803  & -109.47 & R       &  3  &    21.3       &   \\
2018/11/16 & 8438.8     & 3.956  & 3.008  & 4.723  & -109.27 & R       &  3  &    21.3       &   \\
2018/11/18 & 8440.8     & 3.936  & 2.989  & 4.744  & -109.06 & R       &  2  &    21.1       &   \\
2018/11/29 & 8451.7     & 3.828  & 2.917  & 6.471  & -107.90 & R       &  3  &    21.5       &   \\
2018/12/08 & 8461.3     & 3.733  & 2.898  & 9.127  & -106.82 & R       &  3  &    20.3       &   \\
2019/07/28 & 8693.2     & 1.474  & 0.872  & 42.263 & -35.29  & R       &  4  &    17.7       &   \\
2019/08/01 & 8697.2     & 1.451  & 0.740  & 40.927 & -32.28  & R       &  1  &    17.7       &   \\
2019/08/11 & 8706.5     & 1.405  & 0.473  & 28.358 & -24.93  & R       &  24 &    15.9       &   \\
2019/08/17 & 8713.1     & 1.379  & 0.376  & 11.063 & -19.44  & R       &  1  &    15.1       &   \\
2019/08/30 & 8726.3     & 1.346  & 0.589  & 44.078 & -7.96   & R       &  3  &    16.8       &   \\
2020/03/31 & 8940.5     & 3.035  & 2.139  & 9.997  &  97.10  & R       &  2  &    20.2       &   \\
2020/04/03 & 8942.6     & 3.056  & 2.145  & 9.214  &  97.46  & R       &  3  &    20.4       &   \\
2020/04/11 & 8951.4     & 3.147  & 2.195  & 6.830  &  98.93  & R       &  3  &    19.8       &   \\
2020/04/14 & 8954.5     & 3.179  & 2.223  & 6.547  &  99.43  & R       &  2  &    19.4       &   \\
\hline
\enddata
\tablenotetext{a}{Julian Date -2450000.0}
\tablenotetext{b}{Heliocentric, geocentric distance [au]; and phase angle [deg]}
\tablenotetext{c}{True anomaly [deg], the position along orbit; TA at perihelion = 0$^{\circ}$}
\tablenotetext{d}{Magnitude and error through 5$''$ radius aperture; and converted to SDSS r$'$ as described in the text}
\tablecomments{$^{\dag}$MPC data did not include error; assumed $\pm 0.25$ error for light curve calculations}
\end{deluxetable*}

\bibliographystyle{aasjournal}
\bibliography{A2018V3}
\clearpage

\end{document}